\documentclass[english,prl,twocolumn,nofootinbib,superscriptaddress,showpacs]{revtex4}

\makeatletter

\usepackage{babel}
\usepackage{bm}
\makeatother
\begin{document}

\title{The Gravitomagnetic Influence on Gyroscopes and on the Lunar Orbit}

\author{T.~W.~Murphy,~Jr.}
\email[]{tmurphy@physics.ucsd.edu}
\affiliation{University of California, San Diego, 9500 Gilman Drive,
La Jolla, CA 92093-0424}
\author{K.~Nordtvedt}
\affiliation{Northwest Analysis, 118 Sourdough Ridge Road, Bozeman, MT 59715}
\author{S.~G.~Turyshev}
\affiliation{Jet Propulsion Laboratory, California Inst. of Tech.,
4800 Oak Grove Dr., Pasadena, CA 91109}

\begin{abstract}
Gravitomagnetism---a motional coupling of matter analogous to the Lorentz
force in electromagnetism---has observable consequences for any scenario
involving differing mass currents.  Examples include gyroscopes
located near a rotating massive body, and the interaction of two orbiting
bodies.  In the former case, the resulting precession of the gyroscope is
often called ``frame dragging,'' and is the principal measurement sought by
the Gravity Probe-B experiment. The latter case is realized in the earth-moon
system, and the effect has in fact been confirmed via lunar laser ranging
(LLR) to approximately 0.1\% accuracy---better than the anticipated accuracy
of the Gravity-Probe-B result. This paper shows the connnection
between these seemingly disparate phenomena by employing the same
gravitomagnetic term in the equation of motion to obtain both gyroscopic
precession and modification of the lunar orbit.  Since lunar ranging
currently provides a part in a thousand fit to the gravitomagnetic
contributions to the lunar orbit, this feature of post-Newtonian gravity is not
adjustable to fit any anomalous result beyond the 0.1\% level from Gravity
Probe-B without disturbing the existing fit of theory to the 36 years of LLR
data.
\end{abstract}
\maketitle


Part of the post-Newtonian gravitational interaction between two mass
elements, when both are in motion, has been called ``gravitomagnetism,"
in analogy with the magnetic force between moving charges.  The
gravitomagnetic interaction is part of the more general $1/c^2$-order
motional corrections to Newtonian gravity that result from field theories
such as Einstein's general relativity and scalar-tensor generalizations
\cite{gravmag,frame-dragging}.  A total package of velocity-dependent
corrections is required so that the gravitational equation
of motion remains consistent when expressed in different asymptotic inertial
reference frames.  If Lorentz invariance of local gravitational
physics is imposed by empirical constraint, the package of motional
corrections is additionally limited in structure.

The gravitomagnetic interaction of general relativity was first
studied by Lense and Thirring in 1918, and it was shown to produce
both accelerations of and torques on two neighboring rotating bodies.
Others, viewing this phenomenon geometrically, have coined the
interpretive name ``inertial frame dragging" from rotating matter.
It has also been shown that the gravitomagnetic interaction plays a
part in both shaping the lunar orbit at a level (part in a thousand)
readily observable by laser ranging \cite{nord:comp}, and in contributing
to the periastron precession of binary and especially double pulsars
\cite{nord:gravmag}.
\footnote{Although the present analysis uses the equation of motion from
the parameterized post-newtonian (PPN) formulation of long range, metric
gravity, the conclusions we draw are not constrained by this choice.
The phenomenology we explore---the $\bm{v}_i\times (\bm{v}_j\times
\bm{g}_{ij})$ acceleration of Eq.~(\ref{eq:gravterm})---generically
results from modified metric field expansions or even non-metric models
of gravity at the $1/c^2$ post-Newtonian level.  If such an interaction
is used to explain precessional effects in a gyroscope experiment, then
it will also be present to perturb the lunar orbit and binary pulsar
orbits, regardless of its parameterized strength in any particular model.
Any attempt to suppress the gravitomagnetic influence on the lunar
orbit relative to that on a low-orbiting gyroscope by a Yukawa-like
modification to $\bm{g}_{ij}$---whether metric or non-metric---would clash with
strong constraints on the inverse-square nature of $\bm{g}_{ij}$ determined
via satellite and lunar laser ranging.  Eq.~(\ref{eq:gravterm}) is, of
course, dependent on the asymptotic inertial frame in which analysis is
performed, just as are magnetic forces within an electromagnetic system.
Consistent formalisms can be used in any choice of frame to calculate
observables; this permits using the convenient solar system barycentric
frame for our analysis.}

For applications to the analysis of gravitational phenomena, a general
metric tensor field expansion for the gravitational potentials in a broad
class of theories was developed by Will and Nordtvedt \cite{ppn,willbook}.
This parameterized post-Newtonian (PPN) framework yields a gravitomagnetic
contribution to the equation of motion, which in the Lorentz-invariant case
is
\begin{equation}
\bm{a}_{i}=(2+2\gamma)\sum_{j}\frac{\mu_{j}}{c^{2}r_{ij}^{3}}\bm{v}_{i}\times(\bm{v}_{j}\times\bm{r}_{ij}),\label{eq:gravterm}
\end{equation}
where $\bm{v}_{i}$ and $\bm{v}_{j}$ are the velocities of bodies $i$ and
$j$ in the chosen asymptotic inertial coordinate system \cite{nord:comp}.
The vector $\bm{r}_{ij}$, when combined with the fraction
$\mu_{j}/r_{ij}^{3}$ constitutes the Newtonian gravitational acceleration
of mass $i$ toward mass $j$. In geometric language, the PPN factor $\gamma$
quantifies the amount of space curvature produced per unit mass. In general
relativity, $\gamma=1$.  Metric theories allowing preferred inertial frame
effects (absence of local Lorentz-invariance) add the parameter
$\alpha_{1}/4$ to the $(2+2\gamma)$ pre-factor in Eq.~(\ref{eq:gravterm}),
but lunar laser ranging as well as other solar system observations constrain $\alpha_1$ to
be less than $10^{-4}$ \cite{alpha-1}.



We can ask what effect the gravitomagnetic term of
Eq.~(\ref{eq:gravterm}) has on a gyroscope
outside of a rotating spherical mass. 
We define the gravitomagnetic field by
\begin{equation}
\bm{G}_{ij}\equiv(2+2\gamma)\sum_{j}\frac{\mu_{j}}{c^{2}r_{ij}^{3}}(\bm{v}_{j}\times\bm{r}_{ij}),\label{eq:Gij}
\end{equation}
so that $\bm{a}_{i}=\bm{v}_{i}\times\bm{G}_{ij}$ in analogy to the
electromagnetic Lorentz force. Considering a small gyroscope,
the $\bm{G}_{ij}$ vector field is calculated at the gyroscope center,
and will be nearly constant across its body. To
obtain the cumulative effect of mass elements moving within a body
rotating at angular velocity $\Omega$,  the
gravitomagnetic field is integrated over all mass elements, $j$, each with
$d\mu_{j}=G\rho(r_{j})d^{3}\bm{r}_{j},$ where $G$ is Newton's gravitational
constant, and $\rho(r_{j})$ is the mass density at radius $r_{j}$ from the
body center. Adopting a spherical coordinate system aligned with the
rotation axis of the body, we describe the Cartesian vector
$\bm{r}_{j}=r\sin\theta\cos\phi\mathbf{i}+r\sin\theta\sin\phi\mathbf{j}+r\cos\theta\mathbf{k}$,
and the vector to the gyroscope (placed in the $\phi=0$ plane) is
$\bm{r}_{i}=a\sin\psi\mathbf{i}+a\cos\psi\mathbf{k}$, so that
$r_{ij}^{2}=a^{2}+r^{2}-2ar(\sin\psi\sin\theta\cos\phi+\cos\psi\cos\theta)$.
The velocity of mass element $j$ is $\bm{v}_{j}=\Omega
r\sin\theta(-\sin\phi\mathbf{i}+\cos\phi\mathbf{j})$, so that
\begin{eqnarray}
d\bm{G}_{ij} & = & (2+2\gamma)\frac{\Omega r\sin\theta d\mu_{j}}{c^{2}r_{ij}^{3}} \nonumber \\
& \times & \{(r\cos\theta-a\cos\psi)(\cos\phi\mathbf{i} + \sin\phi\mathbf{j})\nonumber \\
& + & (a\sin\psi\cos\phi-r\sin\theta)\mathbf{k}\} .\label{eq:dGij}
\end{eqnarray}

For the special case of a gyroscope situated over the pole
for simplified integration ($\psi=0)$, and recognizing that the $\mathbf{k}$
component of the $\bm{G}_{ij}$ vector will be the only one to
yield a non-zero angular integral, we find that
\begin{eqnarray}
\bm{G}_{ij}(\psi=0) & = & -\frac{(2+2\gamma)G\Omega\mathbf{k}}{c^{2}} \nonumber \\
& \times &\int_{0}^{2\pi}d\phi\int_{-1}^{1}du\int_{0}^{R}dr\frac{\rho(r)r^{4}(1-u^{2})}{(a^{2}+r^{2}-2aru)^{\frac{3}{2}}} \nonumber \\
& = & -\frac{8\pi(2+2\gamma)G\Omega\mathbf{k}}{3c^{2}a^{3}}\int_{0}^{R}dr\rho(r)r^{4}.\label{eq:integration}
\end{eqnarray}
Here we used the identity $u=\cos\theta$, and note that the integral
over $u$ eliminates the $r$-dependence in the denominator.
Recognizing that the moment of inertia of a spherical body is
\begin{equation}
I=\frac{8\pi}{3}\int_{0}^{R}\rho(r)r^{4}dr,\label{eq:inertia}
\end{equation}
we see that
\begin{equation}
\bm{G}_{ij}(\psi=0)=-\frac{(2+2\gamma)GI_{\mathrm{s}}\Omega}{c^{2}a^{3}}\mathbf{k},\label{eq:Gij_at_psi_eq_0}
\end{equation}
where the s-subscript represents the massive rotating body.

Orienting the gyroscope so that its spin axis is along the $\mathbf{i}$
direction: $\bm{\omega}=\omega\mathbf{i}$, 
 the velocity of an element within the
gyroscope is $\bm{v}_{i}=\bm{\omega}\times\bm{r}_{i}=-\omega r\cos\theta\mathbf{j}+\omega r\sin\theta\sin\phi\mathbf{k}$, where
$\bm{r}_{i}$ is the vector position of a mass element within
the gyroscope with respect to its center. Now all  pieces are at hand
to evaluate the acceleration of each mass element within the gyroscope
due to the gravitomagnetic term. The force on each element is
$d\bm{F}_{i}=dm_{i}\bm{a}_{i}=dm_{i}\bm{v}_{i}\times\bm{G}_{ij}$,
and the torque on the gyroscope from this element is then $d\bm{\tau}=\bm{r}_{i}\times d\bm{F}_{i}$.
Combining these steps,
\begin{equation}
d\bm{\tau}=\frac{(2+2\gamma)GI_{\mathrm{s}}\Omega\omega r^{2}\cos\theta dm_{i}}{c^{2}a^{3}}(\cos\theta\mathbf{j}-\sin\theta\sin\phi\mathbf{k}).\label{eq:dtau}
\end{equation}

Integrating this torque over the volume of the gyroscope, 
 the $\mathbf{k}$ component integrates to zero in
the $\phi$ integral, yielding
\begin{eqnarray}
\bm{\tau} & = & \frac{(2+2\gamma)GI_{\mathrm{s}}\Omega\omega}{c^{2}a^{3}}\mathbf{j}\int_{0}^{2\pi}d\phi\int_{-1}^{1}u^{2}du\int_{0}^{R}\rho(r)r^{4}dr \nonumber \\
& = & \frac{(1+\gamma)GI_{\mathrm{s}}\Omega I_{\mathrm{g}}\omega}{c^{2}a^{3}}\mathbf{j},\label{eq:gyro}
\end{eqnarray}
where the integral is seen to be one-half the rotational
inertia of the gyroscope, which is denoted with subscript, g. This torque
will change the angular momentum vector of the gyroscope,
$\bm{L}_{\mathrm{g}}=I_{\mathrm{g}}\bm{\omega}$, such that the angle of the
axis, $\Phi$, precesses at a rate of
$\dot{\Phi}=\left|\bm{\tau}\right|/\left|\bm{L}_{\mathrm{g}}\right|$, so
that
\begin{equation}
\dot{\Phi}=\frac{(1+\gamma)GI_{\mathrm{s}}\Omega}{c^{2}a^{3}}.\label{eq:Phi-dot}
\end{equation}

The direction of precession indicated by Eq.~(\ref{eq:gyro}) is one that takes the angular momentum---originally
in the $\mathbf{i}$ direction only---toward the $\mathbf{j}$ direction,
meaning that the precession has the same sense as the rotation of
the massive body. Had we developed an expression for $\bm{G}_{ij}(\psi=\frac{\pi}{2})$
at the equator of the rotating body, we would have found half the
magnitude of the polar case and in the opposite direction. In general,
the field 
\begin{equation}
\bm{G}_{ij} = -\frac{(1+\gamma)GI_{\mathrm{s}}\Omega}{c^{2}a^{3}}\left[3(\mathbf{k}\cdot\hat{\bm{r}})\hat{\bm{r}}-\mathbf{k}\right] ,\label{eq:Gijfield}
\end{equation}
where $\hat{\bm{r}}$ is the unit radial vector.  In a circular polar orbit with uniform angular rate
in $\psi$, the gravitomagnetic field averages to 
\begin{equation}
\left\langle \bm{G}_{ij}\right\rangle =-\frac{(1+\gamma)GI_{\mathrm{s}}\Omega}{2c^{2}a^{3}}\mathbf{k},\label{eq:Gij_avg}
\end{equation}
 so that the net field shares the same direction as that over the
pole, and therefore the net precession will be in the same sense as
the rotating mass, but at one-fourth the polar rate. Summarizing, 
\begin{equation}
\dot{\Phi}_{\mathrm{polar\ orbit}}=\frac{(1+\gamma)GI_{\mathrm{s}}\Omega}{4c^{2}a^{3}}.\label{eq:phidot_polar}
\end{equation}
 Putting this in another form, where we reduce the rotational inertia
to $I_{\mathrm{s}}=fM_{\mathrm{s}}R^{2}$, where $f=0.33$ for the earth
\cite{stacey}, we have the more convenient form:
\begin{equation}
\dot{\Phi}_{\mathrm{polar\ orbit}}=\frac{(1+\gamma)f}{4}\frac{GM_{\mathrm{s}}}{Rc^{2}}\left(\frac{R}{a}\right)^{3}\Omega.\label{eq:phidot_polar2}
\end{equation}
For Gravity Probe-B (GP-B), in a 640~km altitude polar orbit,
Eq.~(\ref{eq:phidot_polar2}) yields 0.042
arcseconds per year, matching the mission expectation \cite{gpb,will2003}. GP-B anticipates measuring this
precession to $<1$\% accuracy, and perhaps as well as 0.1--0.3\%.  Note
that the gyroscope \emph{spin} was not treated as an intrinsic
property in deriving the gyroscope precession.  Rather, the effect results
from the integrated mass currents of mass elements in 
rotational motion.


In obtaining the effect of the gravitomagnetic term (Eq.~(\ref{eq:gravterm}))
on the lunar orbit relative to Earth, we treat the gravitomagnetic acceleration as a
perturbation about an otherwise circular orbit. We start with an orbit obeying the two-body equation
of motion:
\begin{equation}
\ddot{r}=-\frac{GM}{r^{2}}+r\omega^{2}=a(r)+\frac{l^{2}}{r^{3}},\label{eq:rddot}\end{equation}
where $a(r)$ is the central acceleration, and $l=r^{2}\omega$ is the
(specific) angular momentum. We idealize the unperturbed orbit to have zero
eccentricity and zero inclination, so that the end result is accurate for the
moon at the 10\% level or better.  The deviation, $\delta r$, resulting
from a periodic acceleration perturbation, $\vec{\delta a}$, then obeys
\begin{equation}
\ddot{\delta r}+\omega_0^{2}\delta r=\delta a_{r}+2\omega_0\int^{t}\delta a_{\tau}dt^{\prime},\label{eq:deltar}\end{equation}
where $\omega_0$ is the natural frequency for orbital perturbations,
with $\omega_0^2\approx 3\omega^2-\frac{da}{dr}\approx \omega^2$. The acceleration, $\vec{\delta a}$,
has been decomposed into radial and tangential components, and $t^{\prime}$
is a time variable.

Expressing the triple cross-product in Eq.~(\ref{eq:gravterm}) as
the equivalent dot-product relationship, we find that the gravitomagnetic
acceleration of the moon is\begin{equation}
\bm{a}_{m}\equiv\vec{\delta a}=\frac{\mu_{e}(2+2\gamma)}{c^{2}a^{2}}\left[\hat{\bm{r}}(\bm{v}_{m}\cdot\bm{v}_{e})-\bm{v}_{e}(\bm{v}_{m}\cdot\hat{\bm{r}})\right],\label{eq:gravterm2}\end{equation}
where $\hat{\bm{r}}$ is the unit vector from the earth to the
moon, and $a$ is the earth-moon distance.
Eq.~(\ref{eq:gravterm2}) is re-written as
\begin{equation}
\vec{\delta a}=\frac{(2+2\gamma)GM}{c^{2}a^{2}}\left[\hat{\bm{r}}(V^{2}+\bm{V}\cdot\bm{u})-\bm{V}(\bm{V}\cdot\hat{\bm{r}}+\bm{u}\cdot\hat{\bm{r}})\right],\label{eq:da_gm}\end{equation}
with the earth's
velocity around the sun being $\bm{V}$ and the moon's velocity 
being $\bm{V}+\bm{u}$, where $u$
is approximately thirty times smaller than $V$ in magnitude.

Note that $\bm{u}$ represents to sufficient accuracy the geocentrically-viewed orbital
velocity of the moon around the earth. Thus under the assumption of
a circular orbit (about which we examine the perturbation), $\bm{u}\cdot\hat{\bm{r}}=0$.
Likewise, if we define $\hat{\bm{\tau}}$ to be the tangential
orbit vector at the moon that is perpendicular to $\hat{\bm{r}}$,
$\bm{u}\cdot\hat{\bm{\tau}}=u$. Under the assumption that
the earth is in a circular orbit about the sun, the relationship between
$\bm{V}$ (perpendicular to earth-sun line) and $\hat{\bm{r}}$
and $\hat{\bm{\tau}}$ picks out the synodic phase angle of the moon, $D$.
Specifically, $\bm{V}\cdot\hat{\bm{r}}=-V\sin D$ and $\bm{V}\cdot\hat{\bm{\tau}}=-V\cos D$.
Similarly, $\bm{V}\cdot\bm{u}=-Vu\cos D$. Leaving off the pre-factor from Eq.~(\ref{eq:da_gm}) for now, and dealing only
with the vector math, the radial component is
\begin{eqnarray}
\delta a_{r} & \propto & V^{2}-Vu\cos D-(-V\sin D)^{2} \nonumber \\
& = & \frac{1}{2}V^{2}+\frac{1}{2}V^{2}\cos2D-Vu\cos D,\label{eq:dar_terms}
\end{eqnarray}
and the tangential component is
\begin{equation}
\delta a_{\tau}\propto-(-V\cos D)(-V\sin D)=-\frac{1}{2}V^{2}\sin2D.\label{eq:dat_terms}
\end{equation}
The periodic accelerations consist of two categories: $V^{2}$ terms that have $2D$
angular dependence, and $Vu$ terms that have $D$ angular dependence.
We can treat each separately in solving Eq.~(\ref{eq:deltar}). There
is also a constant term in the expression for $\delta a_{r}$. We
can ignore the constant term since it only acts to rescale the orbit
in a non-periodic way.
We will deal first with the $2D$ terms, then look at the $D$ terms.

First, we integrate the $\delta a_{\tau}$ term. Noting that the rate
at which $D$ advances is $\dot{D}=\omega-\Omega$, the difference
between the lunar orbital frequency and the earth's orbital frequency,
we can construct an arbitrary $2D$ argument as $[2(\omega-\Omega)t^{\prime}+\phi]$,
where $\phi$ is an arbitrary phase depending on the choice of $t^{\prime}=0$.
The integral (without the numerical pre-factor) is then
\begin{equation}
2\omega\int^{t}\sin[2(\omega-\Omega)t^{\prime}+\phi]dt^{\prime}=-\frac{\omega}{\omega-\Omega}\cos2D+\mathrm{const.}\label{eq:atint}
\end{equation}
We have consolidated any initial phase in the integration constant, effectively
defining $t$ so that $D=(\omega-\Omega)t$.
As above, we can ignore the constant term in our periodic analysis.
The differential equation becomes
\begin{eqnarray}
\ddot{\delta r}+\omega^{2}\delta r & = & \frac{(1+\gamma)GM}{a^{2}c^{2}}V^{2}\left[\cos2D+\frac{\omega}{\omega-\Omega}\cos2D\right] \nonumber \\
& = & \frac{(1+\gamma)GM}{a^{2}}\frac{V^{2}}{c^{2}}\frac{2\omega-\Omega}{\omega-\Omega}\cos2D.\label{eq:v2diff}
\end{eqnarray}
The solution, in meters, is then
\begin{equation}
\delta r\approx-(1+\gamma)\frac{V^{2}}{c^{2}}\frac{2-\eta}{1-\eta}\frac{a}{3-8\eta}\cos2D\approx-6.5\cos2D\,\mathrm{m},\label{eq:v2_dr}
\end{equation}
where we have made use of Kepler's relation ($\omega^{2}a^{3}=GM$)
and define $\eta\equiv\Omega/\omega$, ignoring
terms to second order in $\eta$.

The term proportional to $Vu$ has no tangential part, so we immediately
write the differential equation as
\begin{equation}
\ddot{\delta r}+\omega^{2}\delta r=-\frac{(2+2\gamma)GM}{c^{2}a^{2}}Vu\cos D,\label{eq:vu_diff}
\end{equation}
for which the solution is
\begin{eqnarray}
\delta r & = & -\frac{(2+2\gamma)GM}{c^{2}a^{2}}\frac{Vu}{\omega^{2}-(\omega-\Omega)^{2}}\cos D \nonumber \\
& \approx & -(1+\gamma)\frac{Vu}{c^{2}}\frac{\omega}{\Omega}a\cos D\approx-3.4\cos D\,\mathrm{m}.\label{eq:vu_dr}
\end{eqnarray}
But a feedback process produced by the interaction 
of synodic perturbations and the $\cos 2D$ solar tidal distortion of the lunar 
orbit results in an amplification of $\cos D$ terms by the factor
\cite{nord:icarus}
\begin{equation}
Q_{\mathrm{res}}\approx\frac{1-2\eta}{1-7\eta}\approx 1.79,\label{eq:qres}
\end{equation}
so that the corrected range oscillation is
\begin{equation}
\delta r\approx-(1+\gamma)\frac{Vu}{c^{2}}\frac{\omega}{\Omega}\frac{1-2\eta}{1-7\eta}a\cos D\approx-6.1\cos D\,\mathrm{m}.\label{eq:vu_dr_res}
\end{equation}

In summary, the gravitomagnetic perturbations of the lunar orbit are:
\begin{eqnarray}
\delta r_{2D} & = & -6.5\cos2D\,\mathrm{meters} \nonumber \\
\delta r_{D} & = & -6.1\cos D\,\mathrm{meters}. \label{eq:ampls}
\end{eqnarray}


Lunar laser ranging (LLR) has been used for decades to provide a number of
the most precise tests of general relativity, including tests of the weak
and strong equivalence principles, time-rate-of-change of Newton's
gravitational constant, $G$, geodetic precession, among others \cite{jgw:latest}.
Equivalence principle violations would produce a $\cos D$ signal \cite{epsignal},
though no $\cos2D$ signal. Current fits to the archive of LLR data limit
any net deviation of the $\cos D$ term in the lunar orbit to less than
$\approx4$~mm from the orbit prescribed by general relativity \cite{jgw:latest}.
Likewise, the $\cos2D$ term is constrained at roughly the 8~mm level. Thus
barring a chance simultaneous violation of the equivalence
principle \emph{and} gravitomagnetism, the 4~mm constraint translates to a
check on the $\sim$6~m gravitomagnetism amplitude to better than 0.1\%
accuracy. Allowing for such a conspiracy, we must use the 8~mm $\cos2D$
constraint (which is not influenced by equivalence principle violation) to
establish a $\approx0.15$\% verification of the gravitomagnetic phenomenon.
At this time, LLR provides the most precise test of this phenomenon---far
better than the LAGEOS tests of the Lense-Thirring effect \cite{lageos} and
tests from binary pulsars \cite{nord:gravmag}. This result is also likely better than
the ultimate result from the GP-B experiment \cite{gpb}.

A new effort in LLR is poised to deliver order-of-magnitude improvements in
range precision \cite{apollo}, which will translate into tighter constraints on
the $\cos D$ and $\cos2D$ amplitudes in the lunar orbit. Because these are
periodic effects, their accurate determination requires only about a year
of new data collection.  Thus a significantly improved test of this
phenomenon is not far away.


Whether the mass elements in a body are moving commonly---as for the Earth
and the Moon in orbital motion in the solar system---or as mass currents
in the rotational manner of the spinning Earth and gyroscope in GP-B,
the total interaction between bodies must be dominated by a linear-order
integration over the bodies' mass elements in both situations.  Breaking
weak-field superposition would be a radical and ultimately non-viable
modification to gravity theory.  If this linear-order gravitomagnetic
interaction from PPN metric gravity, Eq.~(\ref{eq:gravterm}), is altered
in order to fit any anomalous GP-B observation, then either the $\cos
D$ or $\cos 2D$ amplitudes, or both, in Earth-Moon range as measured
by LLR to half-centimeter accuracy will show anomalies under
this new modeling---establishing a profound empirical clash.  An added
likely consequence of modifying gravitomagnetism would be destroying
the total fit to the binary pulsar 1913+16 data which includes a better
than one percent match to General Relativity's predicted gravitational
radiation-reaction accelerations in that system.  Existing and robust
observations already encumber the gravitomagnetic interaction.

\begin{acknowledgments}
The authors acknowledge useful discussions with Eric Michelsen and Clifford
Will.  The work of SGT was carried out at the Jet Propulsion Laboratory, 
California Institute of Technology, under a contract with the National 
Aeronautics and Space Administration.
\end{acknowledgments}


\begin{thebibliography}{2}
\bibitem{gravmag}Mashhoon, B., arXiV:gr-qc/0311030
\bibitem{frame-dragging}O'Connel, R. F., \emph{Class.
and Quant. Grav.}, \textbf{22}, 3815, (2005)
\bibitem{nord:comp}Nordtvedt, K., arXiV:gr-qc/0301024
\bibitem{nord:gravmag}Nordtvedt, K., \emph{Int. J. of Theo. Phys.}, \textbf{27}, 1395, (1988)
\bibitem{ppn}Will, C. M. \& Nordtvedt, K., \emph{Astrophys. J.}, \textbf{177},
757, (1972)
\bibitem{willbook}Will, C. M., \emph{Theory and Experiment in Gravitational Physics},
Cambridge University Press, New York, (1993)
\bibitem{alpha-1}M\"uller, J., Nordtvedt, K., \& Vokrouhlicky, D., \emph{Phys. Rev. D}, \textbf{54}, R5927, (1996)
\bibitem{stacey}Stacey, F. D., \emph{Physics of the Earth}, Second ed., p. 55,
John Wiley and Sons, New York, (1977)
\bibitem{gpb}Buchman, S., Everitt, C. W. F., et al., 
\emph{Advances in Space Res.}, \textbf{25}, 1177, (2000)
\bibitem{will2003}Will, C. M., \emph{Phys. Rev. D}, \textbf{67}, 062003, (2003)
\bibitem{nord:icarus}Nordtvedt, K., \emph{Icarus}, \textbf{114}, 51, (1995)
\bibitem{jgw:latest}Williams, J. G., Turyshev, S. G., \& Boggs, D. H.,
\emph{Phys.  Rev. Lett.}, \textbf{93}, 261101, (2004)
\bibitem{epsignal}Nordvedt, K., \emph{Phys. Rev.}, \textbf{170}, 1186, (1968)
\bibitem{lageos}Ciufolini, I., Pavlis, E. C., \& Peron, R., \emph{New Astronomy},\textbf{11}, 527, (2006)
\bibitem{apollo}\texttt{physics.ucsd.edu/$\tilde{\ }$tmurphy/apollo/}
\end{thebibliography}
\end{document}